\documentclass[5p]{elsarticle}

\usepackage{lineno,hyperref}
\modulolinenumbers[5]

\journal{Journal of \LaTeX\ Templates}

\begin{document}

\begin{frontmatter}

\title{Dynamics of Network Fluids}

\author{C. S. Dias, N. A. M. Ara\'ujo, and M. M. Telo da Gama}
\address{Departamento de F\'{\i}sica, Faculdade de Ci\^{e}ncias, Universidade de Lisboa, 
    1749-016 Lisboa, Portugal. \\
    Centro de F\'{i}sica Te\'{o}rica e Computacional, Universidade de Lisboa, 
    1749-016 Lisboa, Portugal}

\begin{abstract}
Network fluids are structured fluids consisting of chains and branches.
They are characterized by unusual physical properties, such as, exotic bulk phase diagrams, interfacial roughening 
and wetting transitions, and equilibrium and nonequilibrium gels.
Here, we provide an overview of a selection of their equilibrium and dynamical properties. 
Recent research efforts towards bridging equilibrium and non-equilibrium studies are discussed, as well as several open questions.
\end{abstract}


\end{frontmatter}


\section{Introduction}

Structured fluids are fluids where the particle-particle correlations extend beyond the molecular scale. 
Prototypical examples are suspensions of colloidal particles, where the interactions between 
particles lead to the formation of mesoscopic structures that determine the physical properties 
of the system (e.g., rheological properties). Among the structured fluids are network fluids, where the anisotropic 
particle interactions lead to the formation of dynamical network-like structures consisting of 
chains and branches that are much larger than the individual particles \cite{Witten1990,Witten2004}. Such fluids exhibit exotic phase diagrams, 
including reentrant liquid-vapor or wetting transitions and low density (empty) liquids \cite{Bianchi2006,Russo2011a,Bernardino2012}. 
Examples of network fluids include, suspensions of cross-linked polymers \cite{Witten1988,Stockmayer1943}, and dipolar 
\cite{Klapp2016,Furst2013,Grzelczak2010,Rovigatti2012,DelosSantos2004}, Janus \cite{Kumar2013,Zhang2014a,Jiang2010,Sciortino2010,Munao2015,Hu2012}, 
and patchy \cite{Pawar2010,Bianchi2011,Smallenburg2013a,Song2015,Cho2005,Kim2012,Yi2013,Wang2012,Wang2015,Wang2016,
Kraft2011,Meester2016,Srivastava2014,Vasilyev2015,Vasilyev2013,Groschel2013,Krinninger2016,
Shah2013,Glotzer2004,Glotzer2010,Phillips2014,Kretzschmar2011,Pawar2008,He2012,Lu2013,Shum2010,Lu2008,Manley2005,
Varrato2012,Joshi2016,Rzysko2015,Pizio2014,Sokolowski2014,Kondratowicz2015} particles.

Here, we review some of the equilibrium properties of network fluids and focus on their dynamics. 
We discuss both the bulk \cite{Bianchi2006,Chen2011,Glotzer2007} and interfacial properties \cite{Srivastava2014,Dias2014a,Kumar2013}. 
The manuscript is organized in the following away. The interfacial properties of network fluids are discussed
in Sec.~\ref{sec.interface}, including self-assembly on substrates and at interfaces. The submonolayer regime is the focus of Sec.~\ref{sec.submono}.
In Sec.~\ref{sec.bulk}, the bulk properties are discussed. Experimental and theoretical works on 
the use of programmed annealing cycles to overcome kinetic barriers are discussed  in Sec.~\ref{sec.pathways}. 
Finally, a few concluding remarks are made in Sec.~\ref{sec.final}.

\section{Interfacial properties} \label{sec.interface}

\begin{figure*}[t]
   \begin{center}
    \includegraphics[width=1.8\columnwidth]{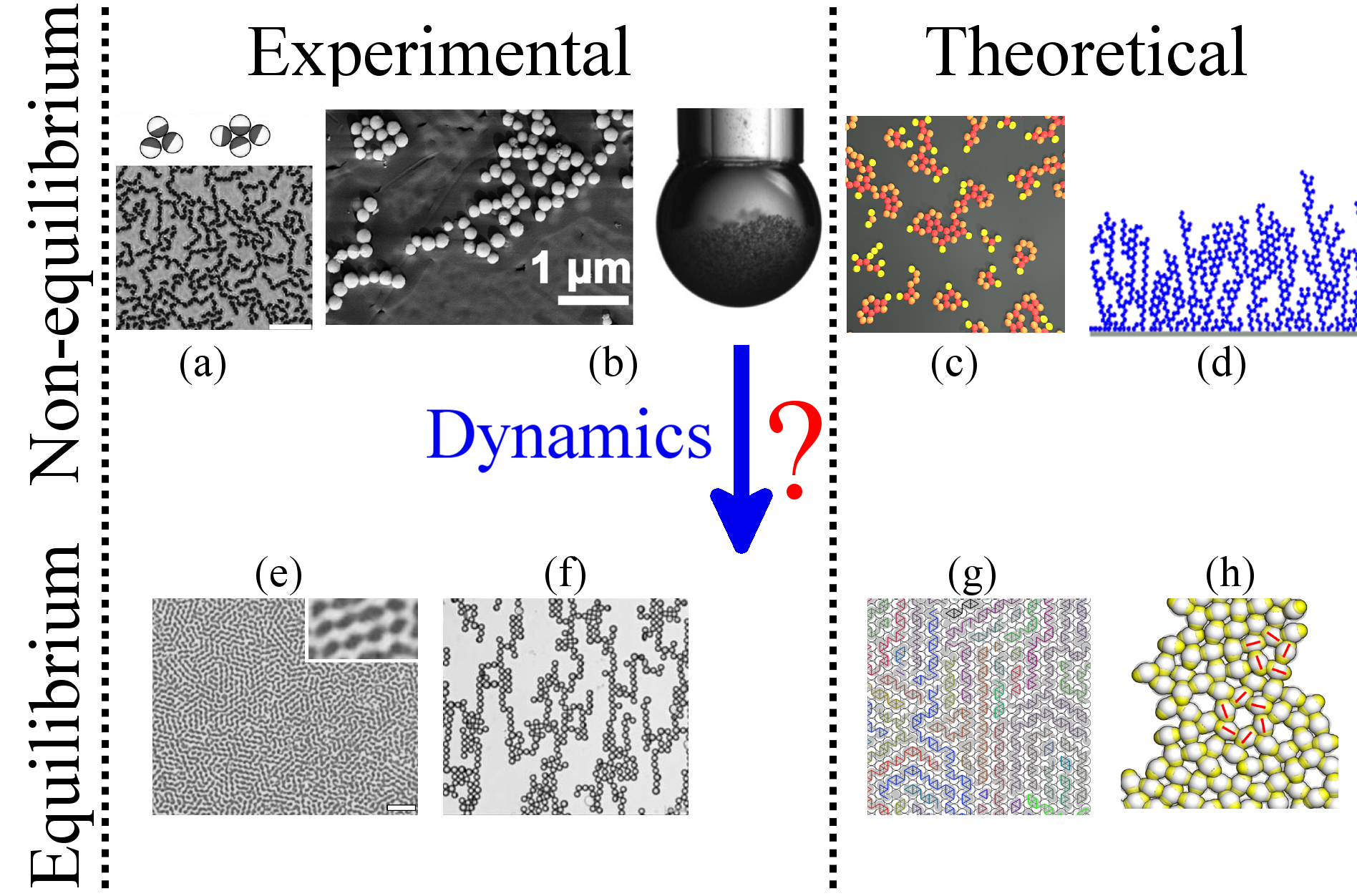} \\
\caption{Self-assembly of network fluids on substrates, or at interfaces, in the limits of irreversible 
(nonequilibrium) and reversible (equilibrium) binding, both from experiments and theory. 
(a) Gold and silica Janus particles on a substrate with a strong particle-particle interaction.
Reproduced from Ref. \cite{Iwashita2013} with permission from the Royal Society of Chemistry; 
(b) Functionalized silica particles at the interface of a water drop. 
Reproduced from Ref. \cite{Fernandez-Rodriguez2015};
(c) Simulations (Langevin dynamics) of three-patch colloidal particles at an attractive substrate. 
Reproduced from Ref. \cite{Dias2016} with permission from the Royal Society of Chemistry;
(d) Kinetic Monte Carlo simulations of three-patch colloidal particles at an interface. 
Reproduced from Ref. \cite{Dias2014} American Physical Society;
(e) Gold and silica Janus particles structure formation. 
Reproduced from Ref. \cite{Iwashita2014} with permission from the Royal Society of Chemistry;
(f) Polystyrene particles with gold patches confined in two dimensions. 
Reproduced from Ref. \cite{Schmidle2013} with permission from the Royal Society of Chemistry;
(g) Pattern obtained from Monte Carlo simulations of Janus particles on a substrate. 
Reproduced from Ref. \cite{Iwashita2014} with permission from the Royal Society of Chemistry;
(h) Monte Carlo simulation of inverse patchy colloidal particles in two dimensions. 
Reproduced from Ref. \cite{Bianchi2013} with permissions from the American Chemical Society.}
  \label{fig.interface_dynamics}
   \end{center}
  \end{figure*}

The equilibrium and non-equilibrium properties of network fluids close to substrates or interfaces 
depend on the anisotropy and strength of the interactions, the temperature, and, for non-equilibrium, the overall dynamics.
In one example, combining equilibrium Monte Carlo (MC) simulations and density functional theory (DFT) for three-patch colloidal 
particles near a hard wall, it was found that a contact region of higher density is formed close to the wall, 
whose maximum density depends on the temperature and bulk density~\cite{Gnan2012}. At high bulk densities, the contact density 
decreases monotonically with the temperature, as an increase in the bond probability favors a decrease in density, 
due to the formation of a low density liquid. At sufficiently low bulk densities, the dependence of the contact density on 
temperature has a minimum at an intermediate temperature, as for very low temperatures, the bond probability approaches unity 
and an ordered, fully connected, quasi-2d structure is formed near the wall that sets a lower bound for the contact density~\cite{Gnan2012}.

The feasibility of the equilibrium structures will depend on the dynamics of self-organization of the colloidal particles. 
As particles start to aggregate, mesoscopic structures are 
formed becoming the relevant units for the dynamics at later times. Investigations of the dynamics are non trivial as they involve 
a hierarchy of processes at different time and length scales \cite{Araujo2017}. The first studies of patchy 
colloidal particles on substrates considered the limit of irreversible bond formation, where 
bond breaking is neglected within the timescale of relevance \cite{Dias2013}. 
Performing kinetic Monte Carlo simulations of a stochastic model, it was shown that the 
structures depend strongly on the mechanism of mass transport \cite{Dias2013b}, number of patches \cite{Dias2015}, and flexibility of the bonds \cite{Araujo2015}. 
For multilayer growth, due to the irreversible nature of the aggregation, 
for diffusive transport the structure is fractal, resembling that of Diffusion Limited Deposition, while 
for ballistic transport, the structure is found to be self-affine \cite{Dias2013b}.  

Recent studies of the irreversible growth of patchy colloidal particles have uncovered phenomena such as roughening transitions,
absorbing phase transitions, and tricriticality \cite{Araujo2015,Dias2014,Dias2014a,Dias2015,Kartha2016,Kartha2016a}. The ballistic aggregation of colloidal particles 
on a substrate is characterized by a growth rate that depends on the relative 
orientation between the patches \cite{Dias2014}. For two-dimensional three-patch particles (disks), it was observed numerically that above and below
a certain value of the opening angle between the patches, the system falls into an absorbing state where, asymptotically, no further 
bonds can be formed. In the limit where two patches are almost overlapping and the 
third one is in the opposite hemisphere of the colloid, the growth of chains 
is promoted, and the transition into an absorbing state is discontinuous. 
In the opposite limit, where the three patches are in the same hemisphere of the particle, the 
transition into the absorbing state is continuous, in the Directed Percolation universality class. 
By adding flexibility to the bonds, the discontinuous transition becomes continuous at 
a tricritical relative orientation between the patches \cite{Araujo2015}. More recently, a lattice version of this model,
for particles with only one patch, was investigated for diffusive \cite{Kartha2016} and ballistic transport \cite{Kartha2016a}.
By decreasing the size of the patch, an absorbing phase transition was found, also in the Directed Percolation
universality class. 

Mixtures have been also considered. There are two types of mixtures: mixtures of particle or patch types.  
The former consists of mixing, at least, two types of particles. A common choice is to consider mixtures of particles of two and 
more than two patches \cite{Bianchi2006,Rovigatti2013}. The idea is to control the branching rate, since
two-patch colloidal particles only form chains and colloidal particles with more patches promote branching. 
A study based on local density approximation of multilayer stacking of 
this type of binary mixtures in a gravitational field has shown stacking diagrams with
different stacking sequences \cite{Geigenfeind2016}. The authors studied the effect of the finite thickness of the 
stacked film on the occurrence of stacks of different layers during sedimentation. A study of the dynamics of sedimentation is still lacking. 

Numerical studies of the dependence on the mechanism of mass transport of binary mixtures towards the substrate have been performed \cite{Dias2013b}. For diffusive 
transport, the density of the multilayer film has a non-monotonic dependence
on the ratio of two- and three-patch particles, with a maximum at an intermediate ratio. 
This non-monotonic behavior is a consequence of the competition between
the density-reducing mechanism of chain formation and an increase of the number of patch-patch bonds per particle,  
by reducing the steric effects. For ballistic transport, the density decreases monotonically with the ratio
and, intriguingly, it is in the same range as the bulk density at the spinodal \cite{Bianchi2006}.

For mixtures of patch types, particles have been considered to have patches of two different 
interaction energies that are distributed with one type in the poles and the other around the equator. In this case, the competition between
branching and chaining depends on the energy ratio between the two types of patches. 
Using a mesoscopic Landau-Safran theory, Bernardino \textit{et al.}~\cite{Bernardino2012} have shown that the 
wetting behavior of these network fluids near substrates
is characterized by a non-monotonic surface tension, two wetting transitions, and a wetting 
transition followed by a drying transition. These interfacial properties can be related 
to the bulk ones for these type of network fluids  \cite{Oleksy2015}. Studies of the dynamics have considered
$2AnB$ patchy colloidal particles ($2$ patches of type $A$ in the poles and $n$ patches of type $B$ along the equator) \cite{Dias2013a}. 
The irreversible aggregation on an attractive substrate of $2A9B$ patchy colloidal particles can lead to non-monotonic behavior
of the density near the substrate as a function of the film thickness depending on the number of patches and energy ratio \cite{Dias2013a}.
For two-dimensional $2A2B$ patchy colloidal particles (disks), 
the interfacial roughening changes from a Kardar-Parisi-Zhang universality class (KPZ) to a KPZ with quenched disorder for B/A energy ratios much lower 
than unity \cite{Dias2014a}.

Most studies of the dynamics of interfaces have considered irreversible bonds and neglected relaxation. However, if one
waits long enough, the interfaces are expected to relax to equilibrium \cite{Bernardino2012}. The bridge between these two
limits is still illusive as it encompasses numerical and theoretical challenges. One step towards closing this gap is the study 
of the submonolayer regime as discussed in the next section.

\section{Submonolayer dynamics} \label{sec.submono}

The submonolayer regime is simpler to study, not only from the numerical point of view, but also from the experimental one, 
as it is possible to follow the dynamics of individual particles
using conventional optical techniques. Experimentally, submonolayer studies have considered, 
for instance, the adsorption on flat and patterned substrates to direct self-assembly into ordered structures \cite{Tian2010,Grzybowski2009}, or
the control of the interface curvature to modify the effective interaction between particles \cite{Fernandez-Rodriguez2015,Garbin2012}.
From the equilibrium point of view, numerical and mean-field phase diagrams have 
shown quantitative agreement in certain limits \cite{Rovigatti2013,Borowko2016,Rzysko2015,Almarza2011}. More recently, 
it was shown that it is possible to tune the submonolayer assembly of
aggregates of heterogeneously charged colloidal particles by confining them between two walls, where one of the walls can be 
electrically charged to promote adsorption \cite{Bianchi2014,Bianchi2013}. 

For mixtures of dipolar colloidal particles, in the presence of an external electric field,
the submonolayer dynamics is characterized by the formation of a spanning aggregate in both directions parallel and transverse to the field.
The dynamics of this aggregate is characterized by a critical slowing down, leading to a power-law decay of the bond correlation function \cite{Schmidle2013,Klapp2016}. 
This type of decay is also observed for three-patch colloidal particles on an attractive substrate, which has been related to a percolation transition \cite{Dias2016}, 
in the limit where patch-patch bonds are practically irreversible. In this limit, the final arrested structures
are significantly different from the thermodynamic ones, and the possible kinetic pathways to overcome them will be discussed 
in Sec.~\ref{sec.pathways}. Also, a study of two systems of three-patch and six-patch colloidal particles, respectively, 
revealed that the dynamics strongly influences 
the initial growth regime \cite{Markova2014}. The authors compared particles with mobile bonds (no fixed orientation over time) and particles
with fixed bonds (fixed orientation over time). They showed that the final cluster is fully connected for
the free bonds by contrast to the fixed bonds case, where only a fraction of the total possible bonds are established.
 
\section{Bulk dynamics}\label{sec.bulk}

The bulk equilibrium phase diagrams of the three models 
of patchy particles presented in the previous section have been studied
in detail \cite{Bianchi2006, Russo2011}. For three-patch colloidal particles and mixtures of two- and three-patch colloidal particles, a low density
(empty) liquid phase is observed \cite{Bianchi2006}. When the interaction between particles has two energy scales, a competition 
between chain and branch formation leads to an unconventional phase 
diagram \cite{Russo2011}, with a reentrant liquid-vapor phase transition. Noteworthy, if the assembly of rings is taken into
account, two critical points may be observed \cite{Rovigatti2013a}.

One of the most exciting features of network fluids is the possibility of obtaining low-density equilibrium gels 
\cite{Sciortino2011,Ruzicka2011,Zaccarelli2007,Zaccarelli2006,Zaccarelli2005,Corezzi2009}, as the percolation line in the phase diagram 
goes above the coexistence line. This allows for reversible gels outside the
phase coexistence region \cite{Sciortino2011}. A relation between temperature, in the reversible
limit and time, in the irreversible limit (chemical bonds) was found \cite{Corezzi2012}.
This is still a very active field of research, where dynamical properties such as the mean square displacement, intermediate scattering functions, 
and van Howe function are measured for suspensions of various types of patchy particles
\cite{Corezzi2008,Corezzi2009,Corezzi2010,DeMichele2006,Elliott2003,Sciortino2009,Sciortino2011,Speedy1995,Zaccarelli2005,Zaccarelli2006,Roldan-Vargas2017,Rovigatti2011}. 

\begin{figure*}[t]
   \begin{center}
    \includegraphics[width=1.8\columnwidth]{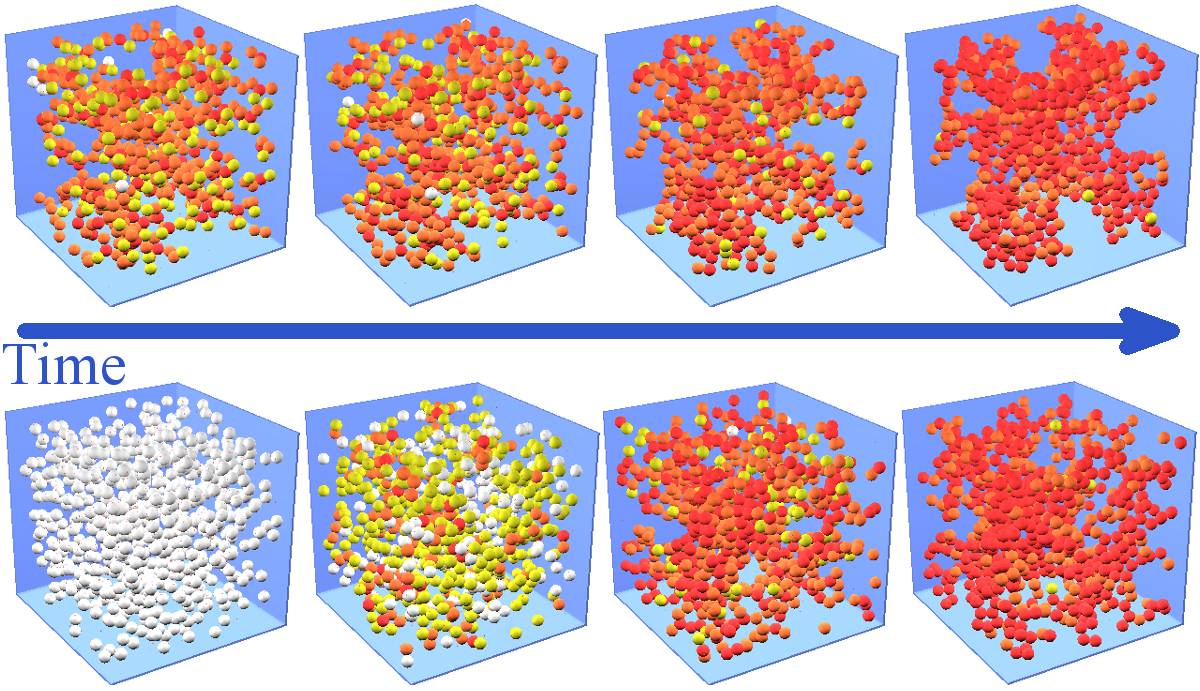} \\
\caption{Snapshots of the time evolution of a network fluid of three-patch colloidal particles starting from
two different initial conditions. Top: Fully connected initial state and Bottom: Unconnected random initial state.}
  \label{fig.bulk_dynamics_snaps}
   \end{center}
  \end{figure*}

All previous results, both for equilibrium and dynamical properties, show that network fluids are very rich systems from the scientific and technological
point of view. Despite that,
most studies have focused on the mechanisms of reversibility of bond formation and chaining versus branching. This is just 
one pair of all mechanisms that play a role in the dynamics. Future studies need to consider how the collective motion of particles inside the
network can influence structure formation, how the rotational diffusion of patches can affect the relaxation dynamics, 
how concerted moves and crowding effects can alter the dynamics, and how the effect of large 
basins on the energy landscape can trap structures into dynamically arrested configurations. An example 
of the latter can be found in Fig.~\ref{fig.bulk_dynamics_snaps}, where independently of the initial conditions, the 
dynamics evolves through a configuration that is significantly different from what is predicted from purely 
thermodynamic arguments \cite{Dias2016a}.

\section{Annealing cycles}\label{sec.pathways}

The idea of performing annealing cycles has been proposed as a kinetic pathway to avoid large kinetic barriers
and access thermodynamic structures of micron size and nanoparticles \cite{Grzelczak2010}. 
The necessity to synthesize structures that are mechanically and thermally stable requires strong interparticle bonds. 
However, the stronger the bonds, the harder it becomes to relax towards thermodynamically stable structures.
To surpass this challenge, the idea of the annealing cycles is to perform protocols of switching on 
and off the bonds to promote the relaxation towards equilibrium. For that end, 
the particles need to be functionalized with different types of molecules that actively react to
external perturbations such as, e.g., light and temperature, or changes in pH, oxidation-reduction, and solubility.

One of the most promising routes is the functionalization of colloidal particles with DNA 
\cite{Leunissen2011,Kern2003,Frenkel2011,Reinhardt2014,Angioletti-Uberti2014,Geerts2010,DiMichele2014}. 
DNA allows a fine tunning of the inter-particle interaction 
strength, by increasing or decreasing the number of nucleotides, and selectivity of the interactions, by changing the sequence of nucleotides. From a dynamical 
point of view, the most important characteristic of DNA molecules is the abrupt melting transition at a well defined temperature. 
This characteristic allows switching on and off bonds
through light (UV/blue) \cite{Bergen2016} or temperature  
\cite{DiMichele2014,Geerts2010,Malinge2016,Mirkin1996,Nykypanchuk2008} cycles, which can even lead to two melting temperatures 
 due to a competition between inter-particle bonds at intermediate temperatures and intra-particle bonds
at low temperature \cite{Angioletti-Uberti2012}. Numerical results suggest that there is an optimal 
annealing frequency at which three-patch colloidal particles on a substrate 
self-assemble into a honeycomb structure \cite{Araujo2017}.

\section{Final remarks}\label{sec.final}

In this overview, we discussed recent findings in the field of network fluids, regarding both equilibrium and non-equilibrium properties. 
The solid framework of equilibrium physics provided the 
tools to study their equilibrium phase diagrams. However, the lack of an equivalent non-equilibrium framework, requires the development of 
new methods and tools to analyze the dynamics. Given the typically strong bonds, of the order of 
several $k_BT$, currently available techniques fail to access the time and length scales at which most relaxation processes occur. 
Thus, most studies of the dynamics have considered irreversible bond formation or very short time periods. 
Clearly, there are still many open questions regarding the dynamics of network fluids.

In order to investigate numerically the dynamics it is mandatory to coarse grain the local interactions, averaging the fast processes and following the rare events. 
A catalog of the relevant processes is needed. Only with such a catalog, 
will it be possible to investigate the feasibility of the thermodynamically stable structures, identify kinetically trapped structures and their stability, 
and develop rules to control the dynamics of the self-organization of network fluids.

\section*{Acknowledgments}

We acknowledge financial support from the Portuguese Foundation for
Science and Technology (FCT) under Contracts nos.
EXCL/FIS-NAN/0083/2012, UID/FIS/00618/ 2013, and IF/00255/2013.

\section*{References}

\bibliography{../Bibtex/SoftMatter}

\end{document}